\documentclass[english,prl, superscriptaddress, longbibliography, twocolumn, notitlepage]{revtex4-1}
\usepackage[T1]{fontenc}
\usepackage[latin9]{inputenc}
\usepackage{array}
\usepackage{amsmath}
\usepackage{graphicx}

\makeatletter

\providecommand{\tabularnewline}{\\}

\usepackage{amsmath}
 
\usepackage[normalem]{ulem}
\usepackage[colorlinks=true,linkcolor=MidnightBlue,urlcolor=MidnightBlue,citecolor=MidnightBlue,anchorcolor=MidnightBlue]{hyperref}\usepackage[dvipsnames]{xcolor}

\usepackage{chngcntr}
\usepackage{xcolor}

\makeatother

\usepackage{babel}
\begin{document}
\title{Age-structured impact of social distancing on the COVID-19 epidemic
in India }
\begin{abstract}
The outbreak of the novel coronavirus, COVID-19, has been declared
a pandemic by the WHO. The structures of social contact critically
determine the spread of the infection and, in the absence of vaccines,
the control of these structures through large-scale social distancing
measures appears to be the most effective means of mitigation. Here
we use an age-structured SIR model with social contact matrices obtained
from surveys and Bayesian imputation to study the progress of the
COVID-19 epidemic in India. The basic reproductive ratio $\mathcal{R}_{0}$
and its time-dependent generalization are computed based on case data,
age distribution and social contact structure. The impact of social
distancing measures - workplace non-attendance, school closure, lockdown
- and their efficacy with duration is then investigated. A three-week
lockdown is found insufficient to prevent a resurgence and, instead,
protocols of sustained lockdown with periodic relaxation are suggested.
Forecasts are provided for the reduction in age-structured morbidity
and mortality as a result of these measures. Our study underlines
the importance of age and social contact structures in assessing the
country-specific impact of mitigatory social distancing.
\end{abstract}
\author{Rajesh Singh}
\email{rs2004@cam.ac.uk}

\affiliation{DAMTP, Centre for Mathematical Sciences, University of Cambridge,
Wilberforce Road, Cambridge CB3 0WA, UK}
\author{R. Adhikari}
\email{ra413@cam.ac.uk}

\affiliation{DAMTP, Centre for Mathematical Sciences, University of Cambridge,
Wilberforce Road, Cambridge CB3 0WA, UK}
\affiliation{The Institute of Mathematical Sciences-HBNI, CIT Campus, Chennai 600113,
India}
\maketitle

\section{Introduction}

The novel coronavirus, COVID-19, originated in Wuhan and has spread
rapidly across the globe. The World Health Organization has declared
it to be a pandemic. In the absence of a vaccine, social distancing
has emerged as the most widely adopted strategy for its mitigation
and control \citep{ferguson2020impact}. The suppression of social
contact in workplaces, schools and other public spheres is the target
of such measures. Since social contacts have a strong assortative
structure in age, the efficacy of these measures is dependent on both
the age structure of the population and the frequency of contacts
between age groups across the population. As these are geographically
specific, equal measures can have unequal outcomes when applied to
regions with significantly differing age and social contact structures.
Quantitative estimates of the impact of these measures in reducing
morbidity, peak infection rates, and excess mortality can be a significant
aid in public-health planning. This requires mathematical models of
disease transmission that resolve age and social contact structures. 

In this paper we present a mathematical model of the spread of the
novel coronavirus that takes into account both the age and social
contact structure \citep{prem2017projecting}. We use it to study
the impact of the most common social distancing measures that have
been initiated to contain the epidemic in India: workplace non-attendance,
school closure, ``janata curfew'' and lockdown, the latter two of
which attempt, respectively, complete cessation of public contact
for brief and extended periods. We emphasise that models that do not
resolve age and social contact structure cannot provide information
on the differential impact of each of these measures. This information
is vital since each of the specific social distancing measures have
widely varying economic costs. Our model allows for the assessment
of the differential impact of social distancing measures. Further,
both morbidity and mortality from the COVID-19 infection have significant
differences across age-groups, with mortality increasing rapidly in
the elderly. It is necessary therefore to estimate not only the total
number of infections but also how this number is distributed across
age groups Our model allows for the assessment of such age-structured
impacts of social distancing measures. 

The remainder of our study is organized as follows. In Section (\ref{sec:Age-and-Contact})
we compare the age and social contact structure of the Indian, Chinese,
and Italian populations. Age distributions are sourced from the Population
Pyramid website \citep{pyramid} and social contact structures from
the state-of-the-art compilation of Prem et. al. \citep{prem2017projecting}\emph{
}obtained from surveys and Bayesian imputation. We show that even
with equal probability of infection on contact, the differences in
age and social contacts in these three countries translate into differences
in the basic reproductive ratio $\mathcal{R}_{0}$. In Section (\ref{sec:Epidemic-without-interventions})
we study the progress of the epidemic in the absence of any mitigation
to provide a baseline to evaluate the effect of mitigation. In Section
(\ref{sec:impact-social-distancing}) we investigate the effect of
social distancing measures and find that the three-week lockdown that
commenced on 25 March 2020 is of insufficient duration to prevent
resurgence. Alternative protocols of sustained lockdown with periodic
relaxation can reduce the infection to levels where social contact
tracing and quarantining may become effective. Estimates of the reduction
in morbidity and mortality due to these measures are provided. We
conclude with a discussion on the possibilities and limitations of
our study. An appendix provide details of our mathematical model and
the social contact structure.

It has been known from retrospective analyses of the 1918--19 pandemic
that delays in introducing social distancing measures are correlated
with excess mortality \citep{Bootsma7588,Hatchett7582}. Our study
confirms the urgency and need for sustained application of mitigatory
social distancing. 
\begin{figure*}[t]
\centering\includegraphics[width=0.98\textwidth]{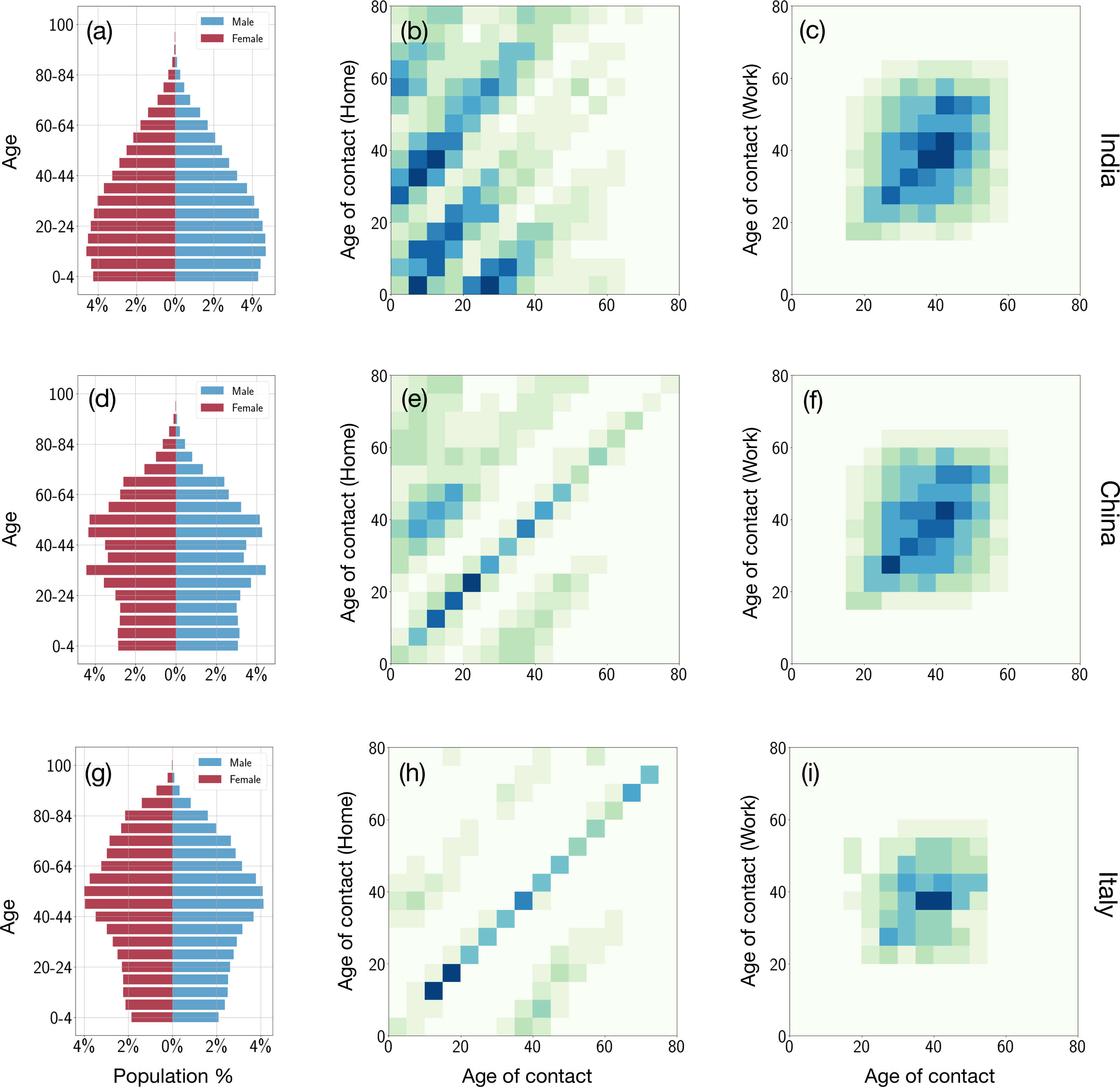}\caption{\textbf{Age and contact structures of the populations of India, China
and Italy}. The first column shows population pyramids by age and
gender. The second and third columns show the contact structures in
households and workplaces with darker colours representing greater
contacts. The diagonal dominance of these matrices shows strong assortative
mixing in all three countries. Significant differences appear in the
off-diagonals. In India, the pentadiagonal character of the household
contacts reflects the prevalance of three-generation households, which
are smaller in China and negligible in Italy. \label{fig:age-contact}}
\end{figure*}

\section{Age and Contact Structures \label{sec:Age-and-Contact}}

In Fig.(\ref{fig:age-contact}) we compare the age and contact structures
of the populations of India, China and Italy. The aim of this comparison
is to highlight their differences and to emphasise the effect these
have on the spread of an infectious disease. Panels (a), (d) and (g)
show the fraction of the population (separated by gender) in five-year
age groups terminating at the age of eighty. The Taj Mahal dome shape
of the Indian age distribution is typical of those undergoing a demographic
transition. The narrower base of both Chinese and Italian populations
is typical of aging populations at or near sub-replacement fertility.
Panels (b), (e) and (h) show the contact between age groups in the
household setting, represented by matrices $C_{ij}^{H}$ where darker
squares indicate larger contacts. As noted in \citep{prem2017projecting},
the features common to all three are the diagonal dominance, reflecting
contact \emph{within} age groups (\emph{i.e. }siblings and partners)
and the prominent off-diagonals, separated by the mean inter-generation
gap, reflecting contacts \emph{between} age groups (\emph{i.e. }children
and parents). The principal difference in India is the presence of
a third dominant diagonal, again separated by the mean inter-generation
gap, reflecting the prevalence of three-generation households. This
quantifies the significant contact between children and grand-parents
and the possibility of substantial of transmission of contagion from
third to first generations. Such contacts are smaller in China and
negligible in Italy. Panels (c), (f) and (i) show the contact $C_{ij}^{W}$
between age groups in the workplace. In contrast to households, the
work contact patterns are more homogeneous across age groups in all
three countries, indicating that the workplace contributes to the
transmission of contagion between age groups that are, otherwise,
largely separated from each other in the household. The boundaries
of these age groups are larger in India and China than in Italy. The
matrices $C_{ij}^{S}$ for schools (shown for India in the appendix)
are strongly assortative, with primary contacts within the school-going
ages and smaller contacts between age groups reflecting student-teacher
interactions. The matrices $C_{ij}^{O}$ for other spheres of contact
(shown for India in the appendix) are strongly assortative, reflecting
the preferential social contact within age groups in this sphere,
but otherwise do not show systematic patterns. In summary, then, in
India the home provides the main channel of transmission between three
generations, the workplace provides the main channel of (largely homogeneous)
transmission between working age groups, the school the main channel
of transmission within children and to a smaller extent between children
and adult teachers, while other spheres of contact, due to the assortative
mixing, contribute to transmission within age groups. 
\begin{table}
\renewcommand{\arraystretch}{1.5} \centering

\begin{tabular}{|>{\centering}p{2cm}|>{\centering}p{4cm}|}
\hline 
Country & Basic reproductive ratio $ $\tabularnewline
\hline 
\hline 
India & $\mathcal{R}_{0}=136\beta$\tabularnewline
\hline 
China & $\mathcal{R}_{0}=117\beta$\tabularnewline
\hline 
Italy & $\mathcal{R}_{0}=119\beta$\tabularnewline
\hline 
\end{tabular}\caption{\textbf{Country-specific basic reproductive ratio} of the age-structured
SIR model for fixed probability of infection on contact $\beta$ and
unit rate of recovery (see text). The difference between countries
is attributed to their differing age and social contact structures.}
\end{table}

Do these differences have a quantitative impact on the transmission
of disease? We answer this affirmatively by comparing the basic reproductive
ratio $\mathcal{R}_{0}$ for each of these populations for an infectious
disease with identical probability of infection on contact $\beta$
and rate of recovery $\gamma$ for the age-structured SIR model described
in Appendix 1. These differences underline the importance of resolving
the age and social contact structure of a population when forecasting
the progress of an infection and the impact of social distancing measures.
With this background, we now turn to our forecast for the progress
of the COVID-19 epidemic in India. 

\section{Epidemic without mitigation\label{sec:Epidemic-without-interventions}}

\begin{figure*}[t]
\centering\includegraphics[width=0.98\textwidth]{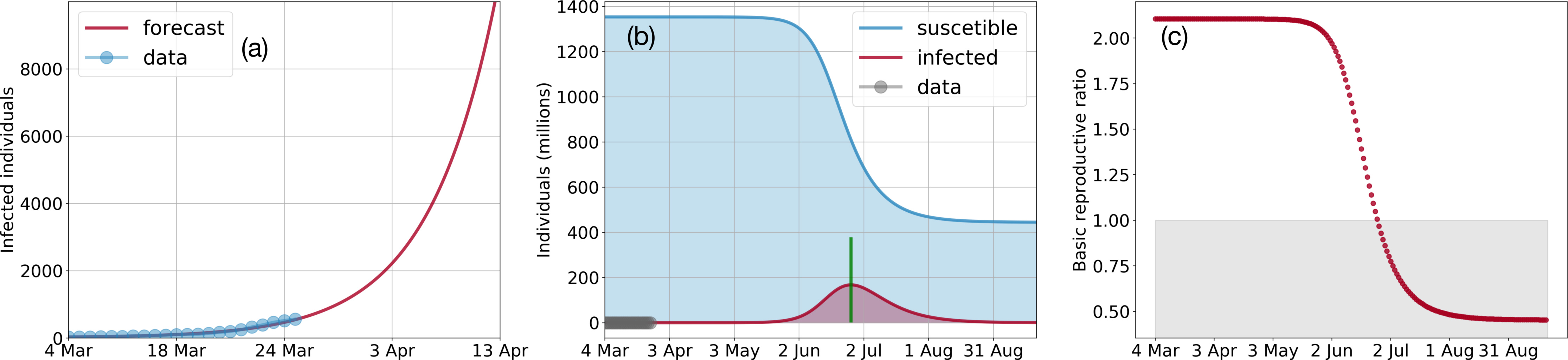}\caption{\textbf{Forecast of the COVID-19 epidemic in India without mitigatory
social distancing.} Panel (a) shows the number of confirmed cases
of till 25th March 2020 (blue circles) and three-week forecast (red
line) from a fit of our model. Panel (b) extends this forecast to
5 months showing the number of infectives (red) and the number of
susceptibles (blue). In the absence of mitigation, an expected 0.9
billion people would be infected in total, with a peak infection of
167 million people in 114 days as indicated by the green bar. Panel
(c) shows the effective basic reproductive ratio $\mathcal{R}_{0}^{\text{eff}}(t)$
as a function of time. This reduces to below unity beyond the peak
infection. This forecast assumes all cases to be symptomatic so $\bar{\alpha}=1$.
The fit parameter $\beta=0.0155$ and we set $\gamma=1/7$.\label{fig:epidemic-curve}}
\end{figure*}

\begin{figure}
\centering\includegraphics[width=0.4\textwidth]{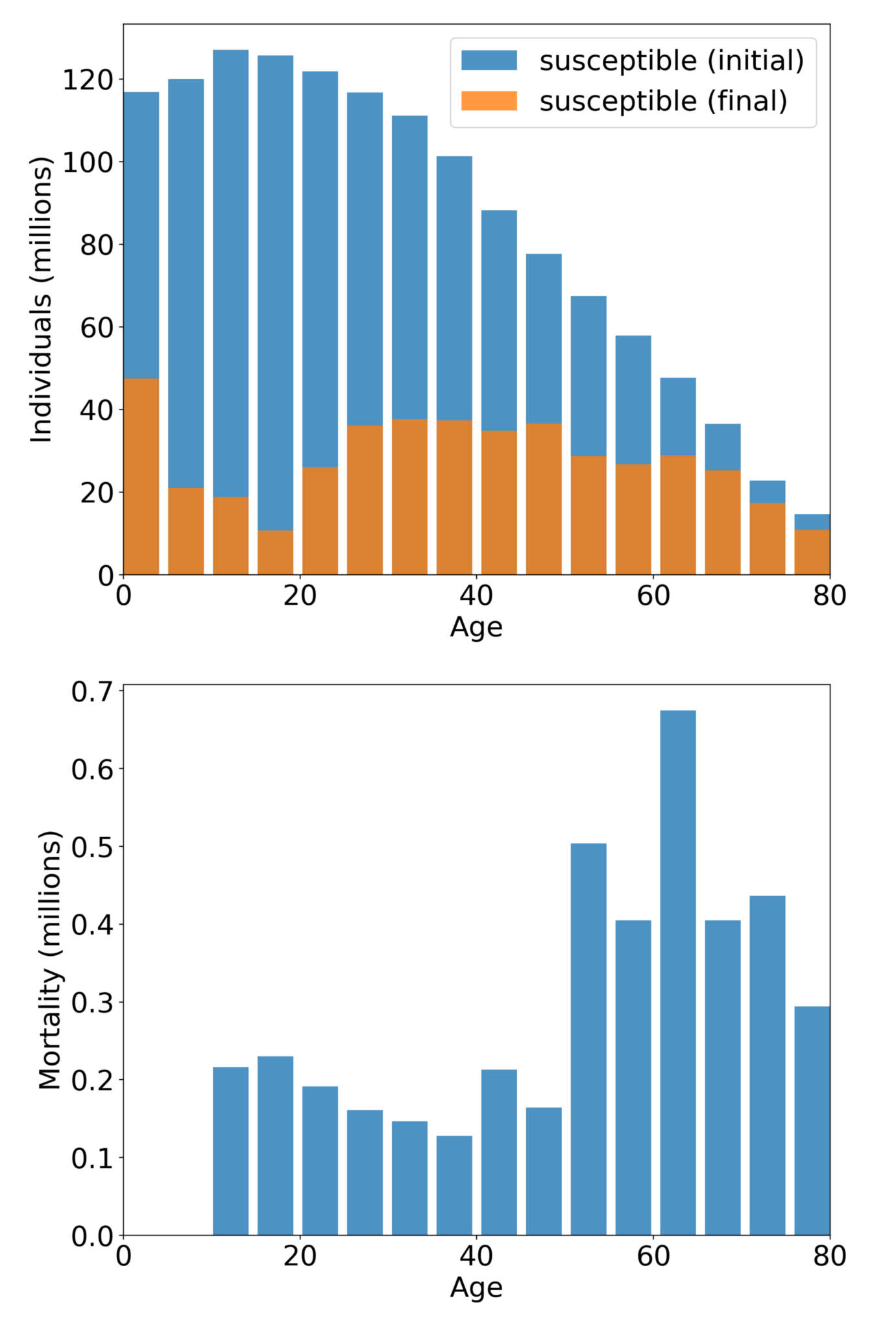}\caption{\textbf{Estimates of morbidity and mortality without mitigatory social
distancing.} The top panel shows the distribution across age groups
of the number of susceptibles at the start of the epidemic (blue bars)
and at the end of the five month forecast (orange bars). Their difference
is the total number infected in that five month period. Greatest infection
is seen amongst the 15-19 year olds and least amongst the 74-79 year
olds. The bottom panel shows the number of mortalities which, due
to the strong age-dependence, is not proportional to the number of
infections. The parameters for these estimates are identical to those
in Fig. (\ref{fig:epidemic-curve}). \label{fig:mortality-morbidity-no-control}}
\end{figure}

We fit our mathematical model, described in detail in Appendix, to
case data to estimate the probability of infection on contact $\beta$.
Though our model allows for infectives to be both asymptomatic and
symptomatic, given the large uncertainty in estimating asymptomatic
cases, we assume all cases to be symptomatic. A possible effect of
this is to underestimate the severity of the outbreak. We then run
the model forward in time to forecast the progress of the epidemic
with results shown in Fig. (\ref{fig:epidemic-curve}). Panel (a)
shows the fit to case data available upto 25th March 2020 and a three-week
forecast, \emph{in the absence} of social distancing measures. The
basic reproductive ratio is $\mathcal{R}_{0}=2.10$. Panel (b) shows
a five month forecast, again, in the absence of social distancing.
The peak infection is reached at the end of June 2020 with in excess
of 150 million infectives. The total number infected is estimated
to be 900 million. Panel (c) shows the time-dependent effective basic
reproductive ratio $\mathcal{R}_{0}^{\text{eff}}(t)$ which gives
the dominant contribution to the linearised growth at any point in
time. This number is greater than unity before peak infection and
smaller than unity beyond peak infection. The serves as a useful measure
of the local rate of change of infectives at any point in time. In
Fig. (\ref{fig:mortality-morbidity-no-control}) we provide estimates
of (a) the morbidity and (b) the excess mortality from the unchecked
spread of the epidemic. The fraction infected across age groups is
the largest for the 15-19 year olds and least amongst the 75-79 year
olds. However, due to the strong age-dependence in death rates, mortality
is amongst the least for the 15-19 year olds and greatest for the
60-64 year olds. We emphasise that these numbers, alarming as they
are, are counterfactuals, as mitigation measures are already in place
of this writing. They do, however, point to the unbearable cost in
human life that must be paid for the any lack of, or delay in, mitigatory
action. 

\section{Impact of social distancing \label{sec:impact-social-distancing}}

We now investigate the impact of social distancing measures on the
unmitigated epidemic. We assume that social distancing in any public
sphere, which in our model is partitioned into workplace, school and
all others, removes all social contacts from that sphere. This, of
course, transfers the weight of these removed contacts to the household,
where people must now be confined. We ignore this in the first instance.
We interpret the lockdown imposed from 25 March 2020 to remove \emph{all}
social contacts other than the household ones. This is an optimistic
interpretation but it does allow us to assess the most favourable
impact of such a measure. The results that follow, then, are an expected
best-case scenarios. Then, the time-dependent social contact matrix
at time $t$ is 
\begin{equation}
C_{ij}(t)=C_{ij}-u(t)(C_{ij}^{W}+C_{ij}^{S}+C_{ij}^{O})
\end{equation}
where $C_{ij}=C_{ij}^{H}+C_{ij}^{W}+C_{ij}^{S}+C_{ij}^{O}$ is the
sum of all social contacts comprising of contributions from the household,
workplace, schools and all others, with obvious superscripts. The
control function, described in Appendix, is constructed to reflect
a social distancing measure that is initiated at $t=t_{\text{on}}$
and suspended at $t=t_{\text{off}}.$ The measure has a lag $t_{w}$
to be effective which we choose to be shorter than a day. The function
varies smoothly from zero to one in the window $t_{\text{on}}-t_{\text{off }}$.
For repeated initiations and suspensions, the control function is
a sum of such terms with times adjusted accordingly. It is possible,
of course, to have differentiated controls which apply distinct social
distancing measures at different times and for different durations.
We do not explore these here as the general setting for such an investigation
would be within the framework of optimal control theory \citep{pontryagin2018mathematical}
with an appropriate cost function. We postpone this to future work.
\begin{figure*}
\centering\includegraphics[width=0.98\textwidth]{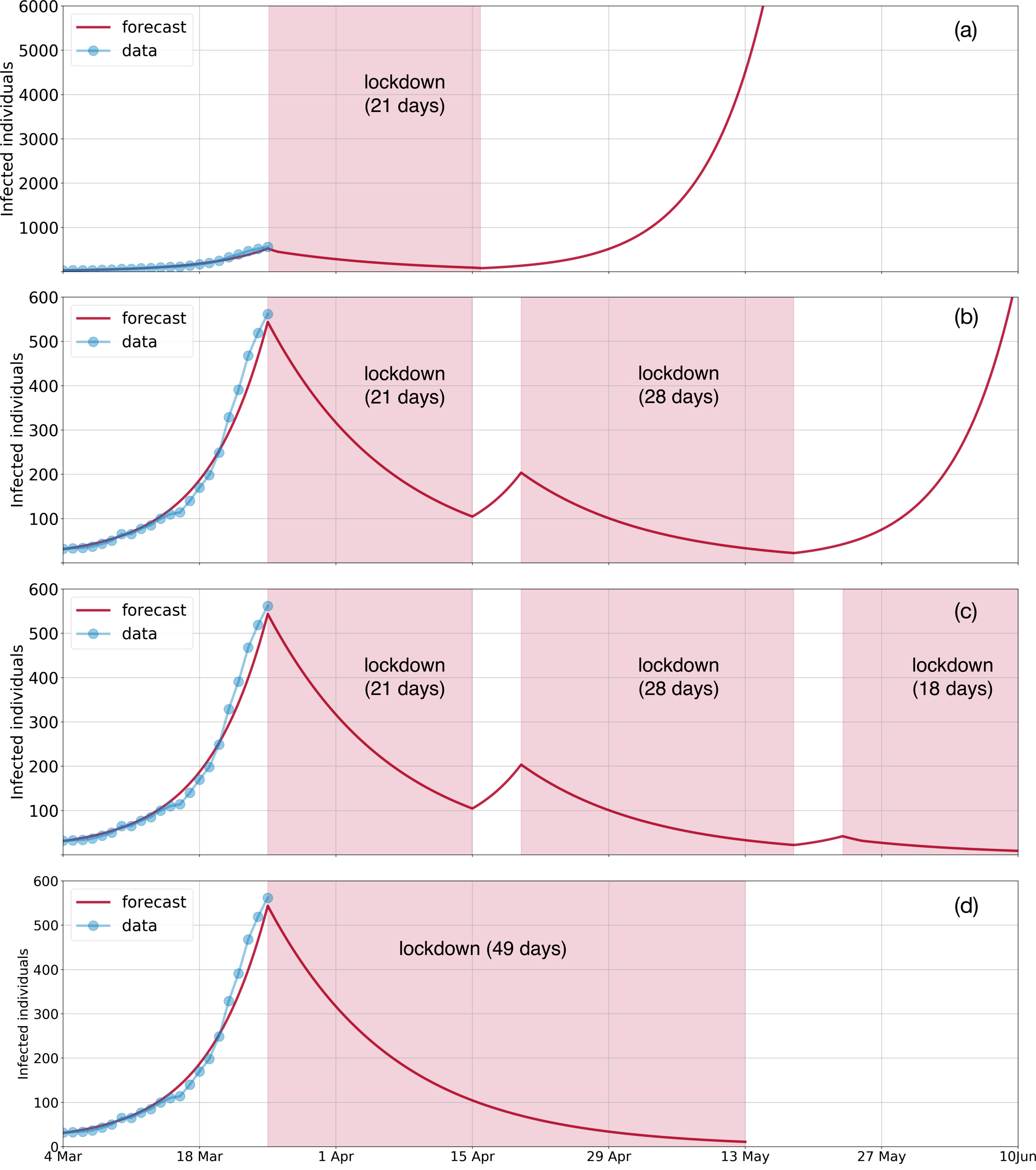}\caption{\textbf{Forecast of the COVID-19 epidemic in India with mitigatory
social distancing. }Each of the four panels shows the variation in
the number of infectives with lockdowns of various durations. The
three-week lockdown starting 25 March does not prevent resurgence
after its suspension as shown in panel (a). Neither does a further
lockdown of 28 days spaced by a 5 day suspension, shown in panel (b).
The protocols in panels (c) and (d), comprising of three lockdowns
with 5 day relaxations and a single 49 day lockdown reduce case numbers
below 10. This forecast is based on all cases being symptomatic so
$\bar{\alpha}=1$. The fit parameter is $\beta=0.0155$ and we set
$\gamma=1/7$. \label{fig:impact-social-distancing}}
\end{figure*}

Our results are show in in the four panels of Fig. (\ref{fig:impact-social-distancing})
for four different control protocols. Panel (a) shows the effect of
the three-week lockdown. While this immediately changes the sign of
the rate of change of infectives, it does not reduce their number
sufficiently to prevent a resurgence at the end of the lockdown period.
Panel (b) shows the effect a suspension of the lockdown by 5 days
followed by a further lockdown of 28 days. This too, does not reduce
the number of infectives sufficiently to prevent resurgence. Panel
(c) shows a protocol of three consecutive lockdowns of 21 days, 28
days and 18 days spaced by 5 days of suspension. This brings the number
of infective below 10 where explicit contact tracing followed by quarantine
may be successful in preventing a resurgence. Panel (d) shows a single
lockdown period to reach the same number of infectives which our model
predicts to be 49 days. 

Table (\ref{tab:mortalit-with-social-distancing}) show the excess
mortality that can be expected for each of the social distancing measures
above. While we emphasise, again, that these are likely to be best-case
scenarios, the substantive message is that of the crucial importance
of rapid and sustained social distancing measures in reducing morbidity
and mortality. 
\begin{table}
\renewcommand{\arraystretch}{1.5} \centering

\begin{tabular}{|>{\centering}p{1.5cm}|>{\centering}p{1.5cm}|>{\centering}p{1.5cm}|>{\centering}p{1.5cm}|>{\centering}p{1.5cm}|}
\hline 
 & Case 1 & Case 2 & Case 3 & Case 4\tabularnewline
\hline 
\hline 
Mortality & 2727 & 11 & 8 & 6\tabularnewline
\hline 
\end{tabular}\caption{\textbf{Estimates of mortality in a 73 day window from 25th March
with mitigatory social distancing.} Cases 1 through 4 correspond,
respectively, to panels (a) through (d) of Fig.\ref{fig:impact-social-distancing}.
The parameters are identical to those in Fig. (\ref{fig:impact-social-distancing}).
\label{tab:mortalit-with-social-distancing}}
\end{table}

\section{Discussion and Conclusion\label{sec:Conclusion}}

We have presented a mathematical model of the spread of infection
in a population that structured by age and social contact between
ages. Since contagion spreads through the structure of social contacts
and the latter varies with age, it is necessary to resolve both these
aspects of a population in any model that attempts to understand and
predict how the modification of the social contact structure through
social distancing impacts the spread of disease. Such models become
useful when reliable estimates of contact structures are available.
We have combined our mathematical model with the state-of-the-art
contact structure compilation of Prem et. al. \citep{prem2017projecting}
and empirical case data available till the 25 March 2020 to assess
the impact of social distancing measures in the spread of the COVID-19
epidemic in India. Our principal conclusion is that the three-week
lockdown will be insufficient. Our model suggests sustained periods
of lockdown with periodic relaxation will reduce the number of cases
to levels where individualised social contact tracing and quarantine
may become feasible. 

Our mathematical model contains both asymptomatic and symptomatic
infectives. Due to the paucity of data on the number of asymptomatic
cases we have chosen to set these to zero. This provides a lower bound
on the number of morbidities and mortalities and the intensity and
duration of the social distancing measures that are required for mitigation.
Extensive testing of the population can provide data on the number
of asymptomatic cases and this, when incorporated into our model,
will provide more accurate estimates of the progress of the epidemic
and the impact of mitigatory social distancing. More generally, there
are uncertainties in all parameters of our model and these would translate
into uncertainties in forecasts and estimates. These uncertainties
can be reduced with better availability of case data and the uncertainties
can be quantified through Bayesian error propagation analysis. The
principal regional differences in India appear to be in the time of
initiation of the infection and for the cases to reach the critical
size where community transmission begins. Though our model is not
spatially resolved, it can be applied region-wise by fitting it to
regional case data.

In closing, we issue the necessary caveats. To quote G. P. Box, ``Since
all models are wrong the scientist must be alert to what is importantly
wrong. It is inappropriate to be concerned about mice when there are
tigers abroad'' \citep{box1976science}. The three components of
our study involve the mathematical model, the sources of data, and
the numerical code. We have provided an explicit description of our
mathematical model, our data sources are referenced, and our numerical
implementation is open-sourced. We take these to be essential desiderata
for modeling to inform policy. The mathematical model of an infection
can aid in qualitative understanding and quantitative prediction but
it should not be used in isolation from other perspectives, including
economic, medical, social and ethical ones. 
\begin{acknowledgments}
R.S. acknowledges the support of a Royal Society-SERB Newton International
Fellowship. RA thanks colleagues at King's College, Cambridge for
their encouragement and forbearance while this work was being completed. 
\end{acknowledgments}

\section*{Appendix A: Mathematical Model \label{sec:Appendix-1}}

\begin{figure*}[t]
\centering\includegraphics[width=0.99\textwidth]{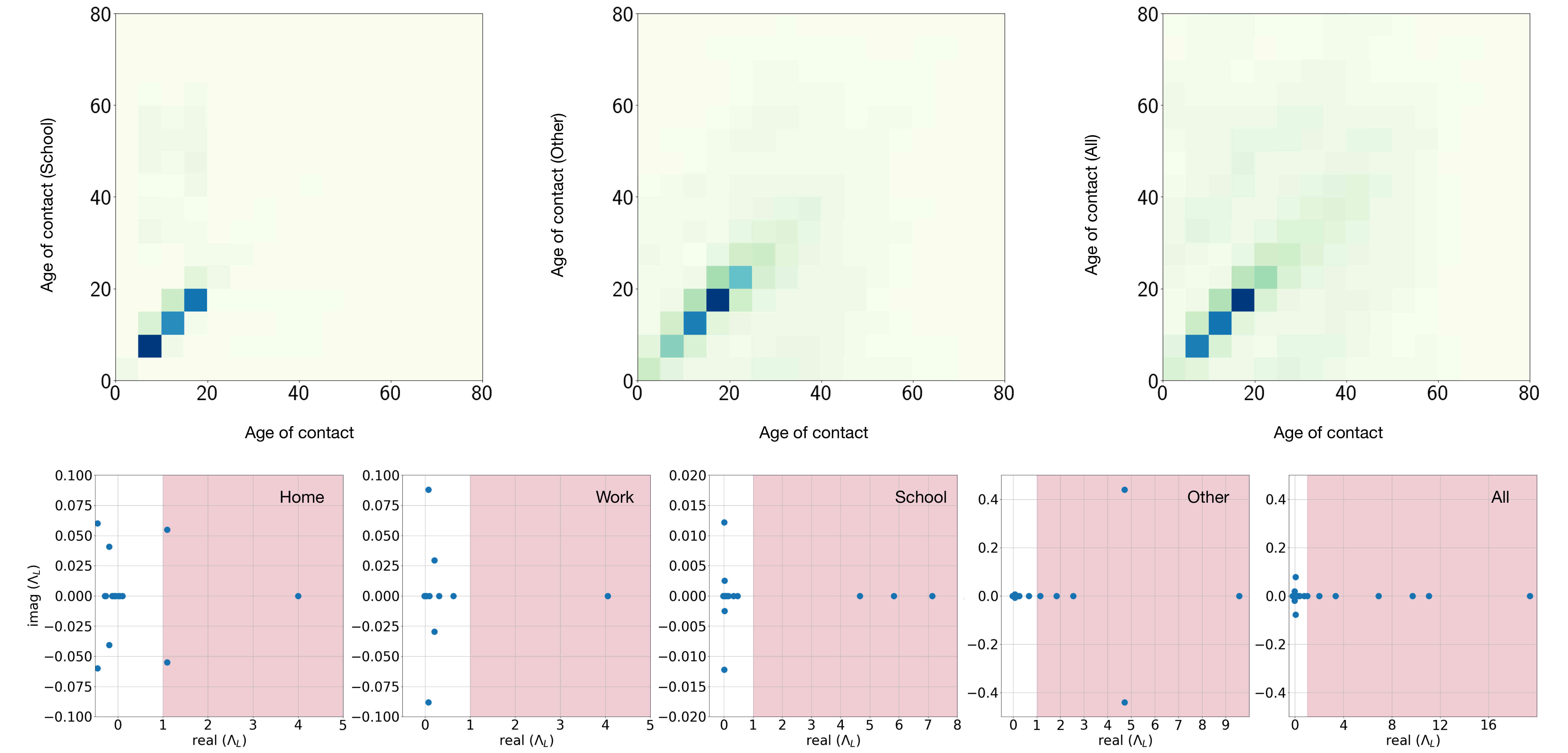}\caption{\textbf{Contact structures of the Indian population and their eigenspectrum.}
The first two figures in the top row show the contact structure in
India in schools and other locations respectively. This completes
the partial list of contact structures in India shown in Fig.\ref{fig:age-contact}.
The third figure shows the sum of all contacts. The large number of
contacts of 15-19 year olds is consistent with their greatest rate
of infection, show in in Fig. (\ref{fig:mortality-morbidity-no-control}).
The second rows show the eigenvalue of the $\boldsymbol{L}$ matrix
for $\bar{\alpha}=\beta=\gamma=1$. The magnitude of the largest eigenvalue
determines the basic reproductive ratio. The eigenanalysis helps estimate
this quantity for each of the structures contributing to social contacts.
\label{fig:epidemic-curve-1}}
\end{figure*}
\emph{ Epidemiological model}: We consider a population aggregated
by age into $M$ groups labelled by $i=1,2,\ldots M$. The population
within age group $i$ is partitioned into susceptibles $S_{i}$, asymptomatic
infectives $I_{i}^{a}$, symptomatic infectives $I_{i}^{s}$ and removed
individuals $R_{i}$. The sum of these is the size of the population
in age group $i$, $N_{i}=S_{i}+I_{i}^{a}+I_{i}^{s}+R_{i}$ \citep{anderson1992infectious,keeling2011modeling,towers2012social,ferguson2006strategies}.
We ignore vital dynamics and the change in age structure on the time
scale of the epidemic. Therefore each $N_{i}$ and, consequently,
the total population size
\[
N=\sum_{i=1}^{M}N_{i}
\]
remain constant in time. We assume that the rate of infection of a
susceptible individual in age group $i$ is \emph{
\begin{equation}
\lambda_{i}(t)=\beta\sum_{j=1}^{M}\left(C_{ij}^{a}\frac{I_{j}^{a}}{N_{j}}+C_{ij}^{s}\frac{I_{j}^{s}}{N_{j}}\right),\quad i,j=1,\ldots M
\end{equation}
where $\beta$ is the probability of infection on contact (assumed
intrinsic to the pathogen) and $C_{ij}^{a}$ and $C_{ij}^{s}$ are,
respectively, the number of contacts between asymptomatic }and \emph{symptomatic}
infectives in age-group $j$ with susceptibles in age-group $i$ (reflecting
the structure of social contacts). We take the age-independent recovery
rate $\gamma$ to be identical for both asymptomatic and symptomatic
individuals whose fractions are, respectively, $\alpha$ and $\bar{\alpha}=1-\alpha$.
With these assumptions the progress of the epidemic is governed by
the age-structured SIR model 
\begin{align}
\dot{S_{i}} & =-\lambda_{i}(t)S_{i},\nonumber \\
\dot{I}_{i}^{a} & =\alpha\lambda_{i}(t)S_{i}-\gamma I_{i}^{a},\label{eq:ageSIR}\\
\dot{I}_{i}^{s} & =\bar{\alpha}\lambda_{i}(t)S_{i}-\gamma I_{i}^{s},\nonumber \\
\dot{R}_{i} & =\gamma(I_{i}^{a}+I_{i}^{s}).\nonumber 
\end{align}
The age structure of the population is specified the proportions $N_{i}/N$
and the contact structure by the matrices $C_{ij}^{a}$ and $C_{ij}^{s}$.
We assume that symptomatic infectives reduce their contacts compared
to asymptomatic infectives and set $C_{ij}^{s}=fC_{ij}^{a}\equiv fC_{ij}$,
where $0\leq f\leq1$ is the proportion by which this self-isolation
takes place. 

\emph{Social contact model}: Partitioning contacts into spheres of
home, workplace, school and all other categories, the contact matrix
can be written as

\begin{equation}
C_{ij}=C_{ij}^{H}+C_{ij}^{W}+C_{ij}^{S}+C_{ij}^{O}.
\end{equation}
For populations of fixed size the contact matrices obey the reciprocity
relation $N_{i}C_{ij}=N_{j}C_{ji}$. 

\emph{Social distancing model}: We model large-scale social distancing
measures by time-dependent controls $u^{W}(t)$, $u^{S}(t)$ and $u^{O}(t)$
imposed on the \emph{non-household} contacts, leading to the time-dependent
contact matrix
\begin{equation}
C_{ij}(t)=C_{ij}^{H}+u^{W}(t)C_{ij}^{W}+u^{S}(t)C_{ij}^{S}+u^{O}(t)C_{ij}^{O}.
\end{equation}
This allows for each one of the possible social distancing measures
to be implemented at different points in and for different durations.
For a lockdown, corresponding to the elimination of all social contacts
other than household ones, a single control function 

\begin{equation}
2u(t)=-\tanh\left(\frac{t-t_{\text{on}}}{t_{w}}\right)+\tanh\left(\frac{t-t_{\text{off}}}{t_{w}}\right)
\end{equation}
is sufficient. Staggered social distancing measures can be constructed
from linear combinations of these controls. 

\emph{Basic reproductive ratio}: We obtain the basic reproductive
ratio by linearising the dynamics about the disease-free fixed point,
where $S_{i}=N_{i}$, and the evolution of infectives is governed
by the $2M\times2M$ linear stability matrix
\begin{equation}
\boldsymbol{J}=\gamma(\boldsymbol{L}-\boldsymbol{1}).
\end{equation}
The $2M\times2M$ \emph{next generation }matrix
\[
\boldsymbol{L=}\begin{pmatrix}\boldsymbol{L}^{aa} & \boldsymbol{L}^{as}\\
\boldsymbol{L}^{sa} & \boldsymbol{L}^{ss}
\end{pmatrix}
\]
consists of the $M\times M$ blocks
\begin{align}
L_{ij}^{aa}=\frac{\alpha\beta}{\gamma}C_{ij}^{a}\frac{N_{i}}{N_{j}},\quad & L_{ij}^{as}=\frac{\alpha\beta}{\gamma}C_{ij}^{s}\frac{N_{i}}{N_{j}},\\
L_{ij}^{sa}=\frac{\bar{\alpha}\beta}{\gamma}C_{ij}^{a}\frac{N_{i}}{N_{j}},\quad & L_{ij}^{ss}=\frac{\bar{\alpha}\beta}{\gamma}C_{ij}^{s}\frac{N_{i}}{N_{j}},
\end{align}
and $\boldsymbol{1}$ is the $2M\times2M$ identity matrix. Collecting
both the asymptomatic and symptomatic infectives in the vector $\boldsymbol{I}=\left(\boldsymbol{I}^{a},\boldsymbol{I}^{s}\right)=(I_{1}^{a},\ldots,I_{M}^{a},I_{1}^{s},\ldots,I_{M}^{s})$,
their dynamics at early times is 
\begin{equation}
\boldsymbol{I}(t)=\exp\left[\gamma(\boldsymbol{L}-\boldsymbol{1})t\right]\cdot\boldsymbol{I}(0).
\end{equation}
Expressing $\boldsymbol{L}$ in its spectral basis of eigenvectors
$\boldsymbol{V}$ and diagonal matrix of eigenvalues $\boldsymbol{\Lambda}=\text{diag}(\Lambda_{1},\ldots,\Lambda_{2M})$
we get $\exp\left[\gamma(\boldsymbol{L}-\boldsymbol{1})t\right]=\boldsymbol{V}\text{diag}[\exp\gamma(\boldsymbol{\Lambda}-\boldsymbol{1})t]\boldsymbol{V}^{-1}$.
For the epidemic to grow, it is sufficient for the spectral radius
of $\boldsymbol{L}$ to be greater than unity. The basic reproductive
ratio is defined to be the spectral radius of $\boldsymbol{L}$ \citep{diekmann2010construction}:

\begin{equation}
\mathcal{R}_{0}\equiv\rho(\boldsymbol{L})=\text{max}\{|\Lambda_{1}|,\ldots,|\Lambda_{2M}|\}.
\end{equation}
If eigenvalue with the largest magnitude is real, the basic reproductive
ratio gives the most dominant contribution
\[
\exp\left[\gamma(\mathcal{R}_{0}-1)t\right]
\]
to the initial growth of the epidemic. This shows that the basic reproductive
ratio depends on \emph{(a) }the probability of infection on contact
$\beta$; (b) the social contact structure encoded in the matrix $C_{ij}$;
(c) the fraction of asymptomatic to symptomatic infectives $\alpha$;
and (d) the fraction by which symptomatic infectives self-isolate
$f$, \emph{i.e. }a combination of pathogen-specific, social, and
individual factors. The linearisation above can be carried out at
any point in time $t$ by making the replacements $N_{i}$ $\longrightarrow$
$S_{i}(t)$ and $C_{ij}\longrightarrow C_{ij}(t)$ in the expression
for $\boldsymbol{L}$, giving the time-dependent stability matrix
$\boldsymbol{L}^{(t)}$. We may define the spectral radius of $\boldsymbol{L}^{(t)}$
as the effective time-dependent basic reproductive ratio 
\begin{equation}
\mathcal{R}_{\text{0}}^{\text{eff}}(t)\equiv\rho(\boldsymbol{L}^{(t)})=\text{max}\{|\Lambda_{1}^{(t)}|,\ldots,|\Lambda_{2M}^{(t)}|\}
\end{equation}
The linearised dynamics gives the number of infectives at time $t+\delta t$
to be
\begin{equation}
\boldsymbol{I}(t+\delta t)=\exp\left[\gamma(\boldsymbol{L}^{(t)}-\boldsymbol{1})\delta t\right]\cdot\boldsymbol{I}(t).
\end{equation}
If the eigenvalue with the largest magnitude is real, the effective
basic reproductive ratio gives the most dominant contribution
\[
\exp\left[\gamma(\mathcal{R}_{0}^{\text{eff}}(t)-1)\delta t\right]
\]
to the short time growth of the epidemic at time $t.$ The number
of infectives will have a negative rate of growth if $\mathcal{R}_{\text{0}}^{\text{eff}}(t)$
is reduced to below unity upon effecting the social distancing measures.
Therefore, $\mathcal{R}_{\text{0}}^{\text{eff}}(t)$ is diagnostic
of the instantaneous efficacy of social distancing measures and can
be computed from the eigenspectrum directly. Assuming controls are
imposed and relaxed periodically and over durations short compared
to intrinsic relaxation times, the progress of the epidemic can be
described as a series of rising and (possibly) falling exponentials
with dominant time scales determined by the spectral radii of $\boldsymbol{L}$
and $\boldsymbol{L}^{(t)}$. To a first approximation, the dynamics
of the rise and fall of infections with the removal and application
of social distancing is governed by a pair of exponentials. The rising
time constant is $\gamma(\mathcal{R}_{0}-1)$ while the falling time
constant is $\gamma(\mathcal{R}_{0}^{\text{eff}}(t_{\text{on}})-1)$.
From Fig. (\ref{fig:epidemic-curve})c it is clear that the spectral
radii are approximately constant away from peak infection times and,
therefore, the time-dependence of the $\mathcal{R}_{0}^{\text{eff}}(t)$
can be neglected to first approximation. 

\emph{Numerical integration}: We choose $M=16$ to correspond to the
$16$ age groups into which the contact matrix data is partitioned.
The $3M=48$ ordinary differential equations are then numerically
integrated using the open source Python library PyRoss which is freely
available on GitHub \citep{pyross}. 

\emph{Data sources}: The data of infected people is obtained from
the website Worldometers \citep{worldometer}. Age distributions are
sourced from the Population Pyramid website \citep{pyramid} and social
contact structures from the state-of-the-art compilation of Prem et.
al. \citep{prem2017projecting}\emph{ }obtained from surveys and Bayesian
imputation.

\end{document}